\documentclass[draft]{agujournal2019}
\usepackage{url} 
\usepackage{lineno}
\usepackage[inline]{trackchanges} 
\usepackage{soul}
\usepackage[autostyle]{csquotes}
\draftfalse

\journalname{JGR: Planets}

\begin{document}

%
%


\title{Effect of lunar rotation on antipodal ejecta deposition}

%
%




\authors{S. Wakita\affil{1}, B. C. Johnson\affil{1,2}}

\affiliation{1}{Department of Earth, Atmospheric, and Planetary Sciences, Purdue University, West Lafayette, IN, USA}
\affiliation{2}{Department of Physics and Astronomy, Purdue University, West Lafayette, IN, USA}




\correspondingauthor{Shigeru Wakita}{swakita@purdue.edu}



\begin{keypoints}
\item We track the location of impactor material deposited around the antipode including the effect of lunar rotation.
\item Our work shows that rotation of the Moon displaces antipodal ejecta, especially for an impact at the equator. 
\item The effect of rotation should be considered when interpreting lunar magnetic anomalies located antipodal to basins. 
\end{keypoints}

%
%

%
%

\begin{abstract}
The Moon's crust exhibits strong magnetic anomalies that likely record ancient magnetic fields produced by a core dynamo. 
Lunar materials have low magnetic susceptibility, but impactor material with high susceptibility may record the Moon's ancient dynamo. 
Some magnetic anomalies are antipodal to lunar basins and it is hypothesized that the impact ejecta accumulates at the basin antipode and records the ancient lunar magnetic field.
Previous impact simulations, which ignored the rotation of the Moon, support this hypothesis. 
Here, we reevaluate the deposition of antipodal ejecta by considering the rotation of the Moon. 
Assuming that the rotational period of the Moon is the same as the current day, we find that the rotation of the Moon does not strongly affect the deposition of the antipodal ejecta if impacts occur near the poles. 
In contrast, rotation has a considerable effect on antipodal deposits for impacts occurring near the equator. 
An eastward impact at a 45 degree incidence angle results in almost no antipodal ejecta, however, an eastward impact with a 30 degree incidence angle accumulates more antipodal ejecta than the non-rotating case.  
It is hard to determine the impactor's direction and location, for basins without an ellipsoidal shape and/or an asymmetric ejecta blanket. 
Nevertheless, our work demonstrates that care must be taken in interpreting the magnetic anomalies antipodal to lunar basins and that the effect of rotation should not be ignored.
\end{abstract}

\section*{Plain Language Summary}
Lunar spacecraft have detected strong localized magnetic fields in the lunar crust, possibly the result of an ancient core dynamo.
However, lunar materials are difficult to magnetize.
Crater-forming asteroids are likely to contribute to the recording of the ancient lunar magnetic field. 
Some observed magnetic fields are located at the opposite side of the Moon or antipodal to lunar basins. 
Several studies have hypothesized that asteroidal materials accumulated at the basin antipode. 
Previous impact simulations showed that ejected materials from oblique impacts gather at the impact antipode, but that work ignored the rotation of the Moon. 
Our work examines the deposition of impact ejecta at the antipode of lunar basins, considering lunar rotation. 
Lunar rotation has less effect on antipodal ejecta deposition if the impact occurs near the pole. 
For the impact near the equator, antipodal ejecta thickness varies based on the impact angle; increasing, decreasing, or non-existent.
The distribution of ejecta at the antipode of the lunar craters also depends on the impact angle and the Moon's rotation period. 
Although decoding the details of impactors, such as direction and angle, is difficult, this work suggests that lunar rotation should be included in understanding the lunar crustal magnetic fields.

%
%

%


%
%
%
%

\section{Introduction} \label{sec:introduction}
The Moon's crust exhibits strong magnetic anomalies despite the absence of a current dynamo magnetic field. 
Thus, localized magnetic anomalies provide evidence of ancient magnetic fields \cite{Hood:2008,Tsunakawa:2015}.
Due to the low susceptibility of lunar material, exogenous material is thought to be responsible for crustal magnetic anomalies \cite{Strangway:1973,Moore:1974,Wieczorek:2012}.
For example, magnetic anomalies within and around basins could be remnants of the basin-forming impactor \cite{Wieczorek:2012}.
Some magnetic anomalies are located antipodal to lunar basins, such as Imbrium, Serenitatis, Orientale, and Crisium \cite{Hood:2008,Lin:1988,Mitchell:2008,Hood:2021}. 
Impact ejecta accumulated at the basin antipode may produce these magnetic anomalies \cite{Hood:2008,Wakita:2021}.

Basin-forming impacts excavate and distribute ejecta over the Moon. 
Pressure remanent magnetization upon ejecta emplacement is unlikely because ejecta experience low-shock pressures \cite{Tikoo:2015,Garrick-Bethell:2020}. 
In contrast, thermoremanent magnetization is possible as the impact material could have higher susceptibility than the lunar crust \cite{Wieczorek:2012} and remain above the Curie temperature until after emplacement \cite{Wakita:2021}. 
Assuming a chondritic meteorite obliquely striking the Moon, previous work indicated that the accumulation of impactor material is capable of recording the ancient magnetic field \cite{Wakita:2021}. 
\citeA{Citron:2025} also investigates lunar antipodal ejecta, however, both works assumed no rotation of the Moon.
Rotation will effect ejecta emplacement especially for late arriving material. 
This in turn can effect the strength and size of antipodal magnetic anomalies. 

The effect of rotation on ejecta deposition on planetary bodies has been previously investigated. 
\citeA{Dobrovolskis:1981} analytically examined the trajectories of ejecta on rotating planetary bodies that modify the usual trajectories on non-rotating bodies.
Distribution of ejecta on rotating bodies has been investigated on the Earth \cite{Wrobel:2003}, Mars \cite{Wrobel:2004}, and Moon \cite{Wieczorek:2001}. 
Using an analytical approach, previous work proposed that the antipodal ejecta from Serenitatis basin could be responsible for the ejecta deposition at Imbrium and South Pole-Aitken basins, observed as mounds and thorium anomalies \cite{Wieczorek:2001}. 
The main limitation of these studies is the use of simple analytical models of ejection of material by the impact (i.e., ejecta scaling model).
Here, we reevaluate the deposition of antipodal ejecta by considering the rotation of the Moon, using results from high resolution three-dimensional numerical simulations of basin forming impacts \cite{Wakita:2021}. 
The use of results from high resolution simulations allows us to track the trajectory of impactor materials which are more likely to record the Moon’s magnetic field \cite<e.g.,>{Wieczorek:2012}.
By varying the impactor direction and location (i.e., poles or equator), we explore the effect of lunar rotation on antipodal ejecta deposition of both impactor and target material. 

\section{Methods} \label{sec:methods}
We examine the effect of lunar rotation on the distribution of antipodal ejecta. 
We use previous results from impact simulation of 100 km-diameter-impactors striking the Moon at 12 km/s at incidence angles of 30, 45, and 60 degrees \cite{Wakita:2021}. 
Note that 12 km/s is lower than the mean impact velocity on the Moon \cite<17.4 km/s,>{Yue:2013}. 
We select this velocity because a higher impact velocity produces an abundance of molten ejecta which is unlikely to generate magnetic anomalies \cite{Wakita:2021}. 
The previous work examined those impacts as a dunite impactor striking a flat dunite target \cite{Benz:1989} with 1 km resolution. 
They emplaced a Lagrangian tracer particle in each computational cell before the impact (14279170 tracer particles in total).
Tracer particles record their position and velocity during the impact simulations. 
They took ejecta properties when they reached 50 km from the target, which happens earlier in crater formation.
Using analytical equations of \citeA{Dobrovolskis:1981}, they calculate the trajectories and deposition of tracer particles on the non-rotating Moon \cite{Wakita:2021}. 
For more details of their models, please see \citeA{Wakita:2021}.
The distributions on the antipodal hemisphere are shown in Figures \ref{fig:non45degree}, \ref{fig:non30degree} and \ref{fig:non60degree} for impact angle of 45$^\circ$, 30$^\circ$, and 60$^\circ$, respectively.
Most of ejecta that land near the impact location (i.e., right halves of the circle representing location between the impact location and the antipode) have a flight time of less than 10 hours.
In contrast, those landing over the antipode have a longer flight time.
Note that the flight time of ejecta on the hemisphere near the impact point is shorter than 1 hour (not shown here). 
The effect of rotation is minimal (up to about 1 degree offset), and we omit their analysis since it is not the focus of this paper.
We calculate the displacement of ejecta according to their flight time and the rotational period of the Moon. 
Although the current lunar rotational period is $\sim$ 27 days, the ancient Moon rotated faster \cite{Dwyer:2011,Daher:2021}.
While we use the rotational period as 27 days as a fiducial case, we also consider 17 days, which is the estimated rotational period 4.5 Ga \cite{Daher:2021}.
For the impact direction and location, we examine four extreme cases: a southward impact at the north pole, a northward impact at the south pole, a westward impact at the equator, and an eastward impact at the equator. 
Because we expect that an impact at the pole will result in no displacement of antipodal ejecta deposition even considering lunar rotation. 
For an impact at the equator, it will maximize the rotational effect on the antipodal deposition. 
Thus, our four extreme cases will cover the results of other impacts occurring at latitudes other than the pole or the equator.

\section{Results} \label{sec:results}
\begin{figure}
\noindent\includegraphics[width=\textwidth]{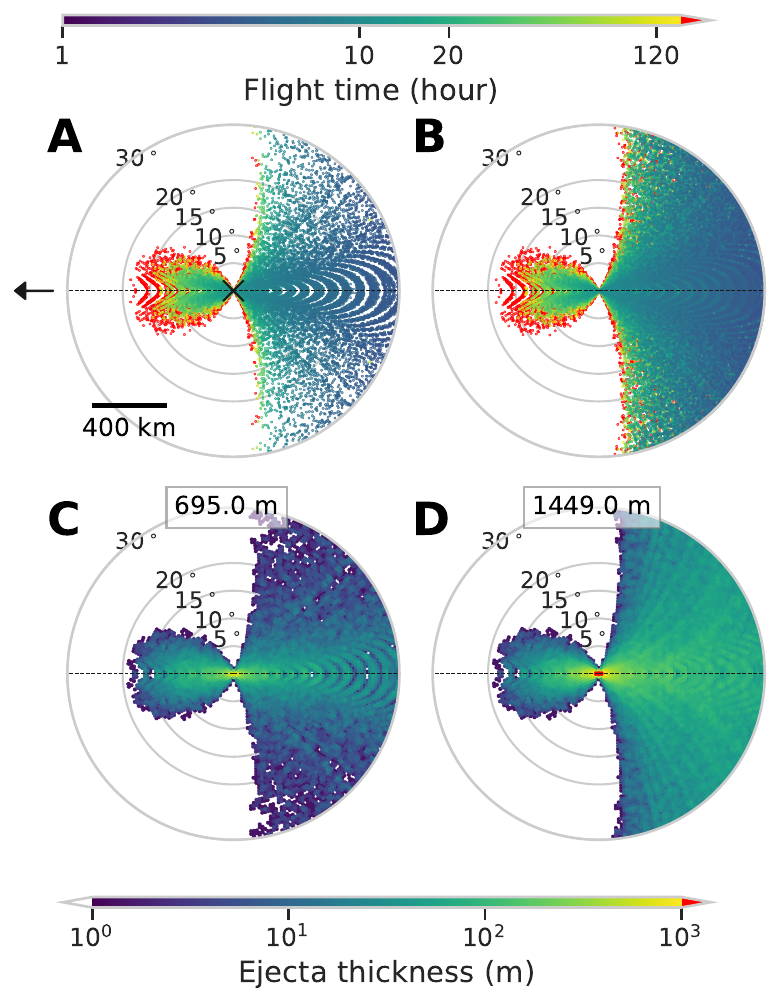}
\caption{
Distribution of impactor ejecta at antipodal hemisphere.
Panel A and B illustrate the flight time, and C and D are the thickness of impactor ejecta from a 100 km-diameter-impactor at 12 km/s and 45$^\circ$.
Note that this is the non-rotating case.
Left panels show impactor ejecta and right panels show total (impactor and target) ejecta. 
The value in the box of C and D indicates the thickness at the antipode.
Panel A also depicts the black cross at the center corresponding to the antipode of the impact site. 
The arrow indicates the downrange direction of the impactor. 
While each point in Panel A and B represents each tracer particle, each point in Panel C and D corresponds to the surface area of 1$^\circ$, in which we derive its thickness from the volume of ejecta \cite<see also>{Wakita:2021}.
Note that the numerical resolution (i.e., 1 km) causes the arc-shaped distribution. 
}
\label{fig:non45degree}
\end{figure}

\begin{figure}
\noindent\includegraphics[width=\textwidth]{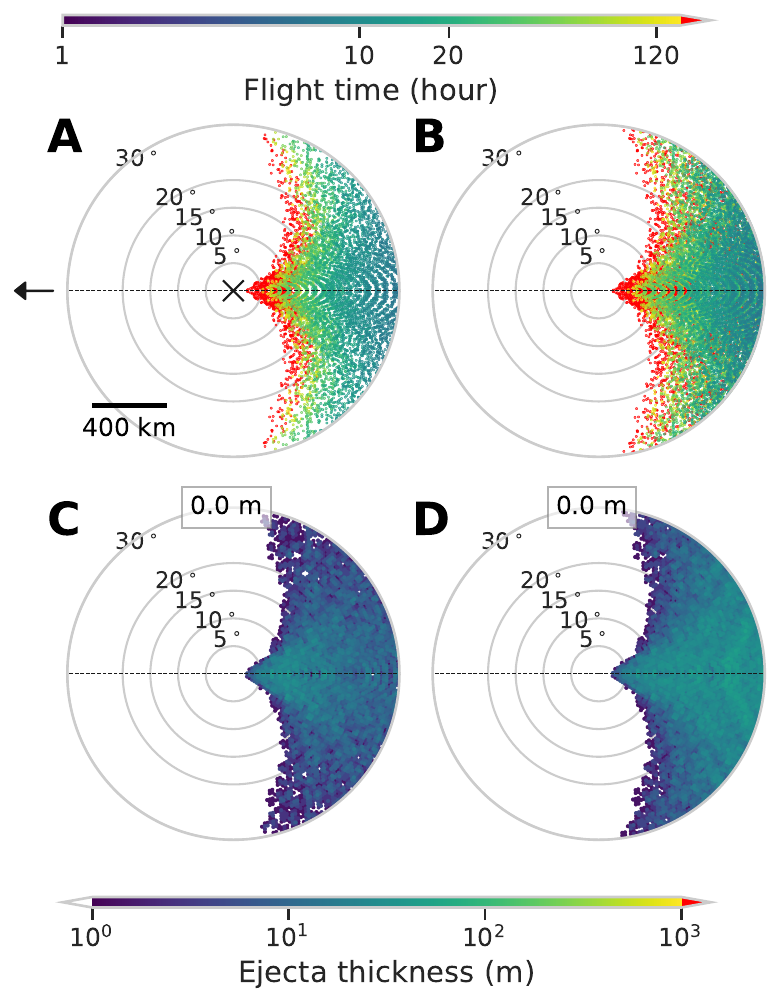}
\caption{
Same viewing as Figures \ref{fig:non45degree}, but the impact angle is 60$^\circ$.
Note that this is the non-rotating case.
}
\label{fig:non60degree}
\end{figure}

\begin{figure}
\noindent\includegraphics[width=0.6\textwidth]{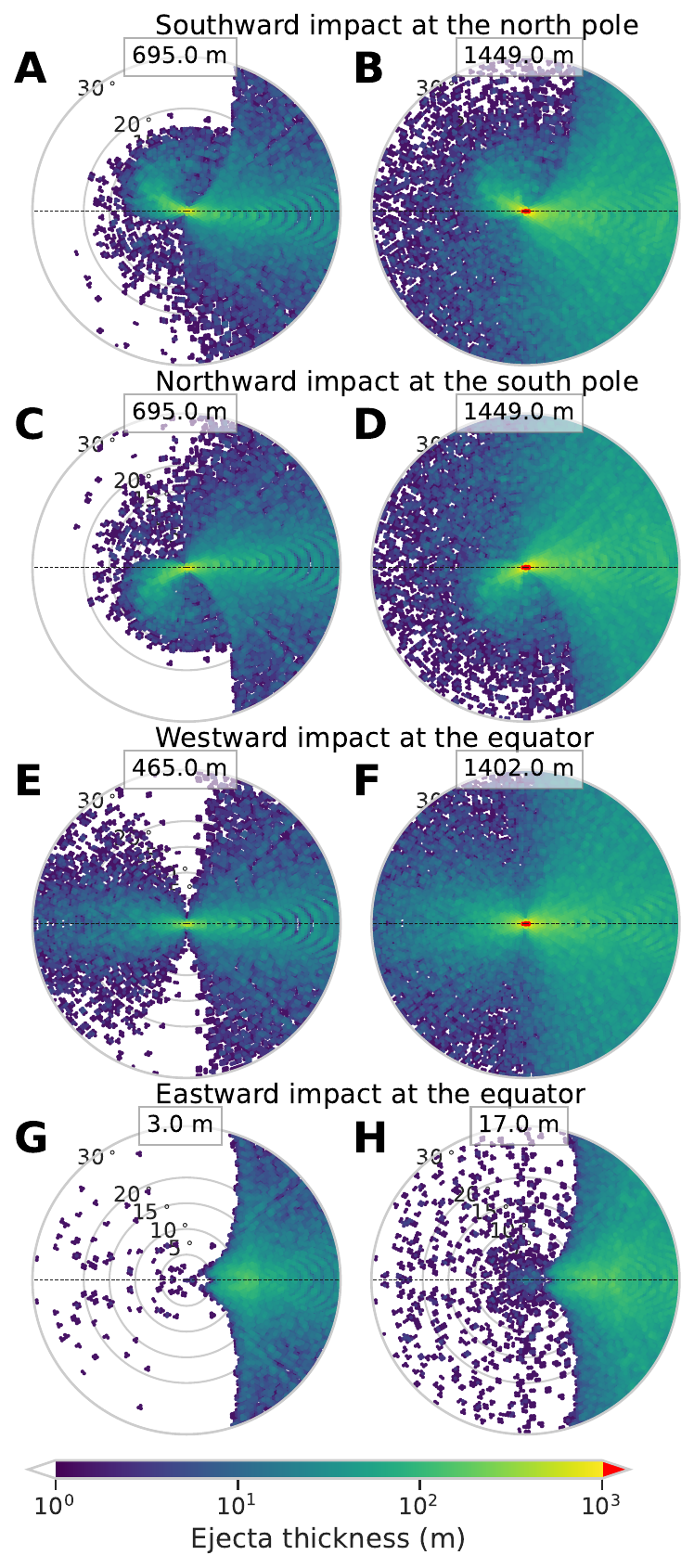}
\caption{
Distribution of impactor and total ejecta at antipodal hemisphere.
Left panels show impactor ejecta and right panels show total (impactor and target) ejecta. 
Note that the numerical resolution (i.e., 1 km) causes the arc-shaped distribution (see A and B).
Panels correspond to a different impactor direction and impact site: 
A) and B) a southward impact at the north pole,
C) and D) a northward impact at the south pole,
E) and F) a westward impact at the equator,
and G) and H) an eastward impact at the equator, respectively.
Note that these are shown in an antipode focused coordinate system (see Section \ref{sec:results}).
The center is fixed as the antipode of the impact site and tick marks indicate distance from the antipode in degrees, not latitude nor longitude on the Moon.
The value in the box indicates the thickness at the antipode. 
The distribution along the impact direction of C--F (black dashed line) is shown in Figure \ref{fig:rot_along}.
We note that the rotational period for these cases is 27 days.
}
\label{fig:rot}
\end{figure}

The rotation of the Moon does not strongly affect the deposition of the antipodal ejecta when impacts occur at the poles.
Figure \ref{fig:rot} illustrates the ejecta thickness at the antipodal hemisphere in the case of a 45$^\circ$ impact. 
To focus on the distribution of antipodal ejecta, we keep displaying it at the antipodal hemisphere throughout this work. 
While the Moon rotates at the pole, the center of each panel is fixed as the antipode of the impact site.
Using this antipodal focused coordinate system makes it easier to compare simulations of non-rotating and rotating cases. 
We show the ejecta from the impactor on left panels and the total ejecta on right panels. 
We note that the impactor ejecta have a high enough magnetic susceptibility to record the magnetic field (see Section \ref{sec:introduction}), thus we mainly focus on their results.
The ejecta distribution reflects the direction of the lunar rotation, and the results of impacts at the north pole and the south pole are mirrored (Figure \ref{fig:rot}A and \ref{fig:rot}C). 
The ejecta emplaced past the antipode, which has a longer flight time (Figure \ref{fig:non45degree}A), is displaced more than ejecta landing before the antipode (i.e., right halves of the circle Figure \ref{fig:rot}A and \ref{fig:rot}C), which has a shorter flight time.
Regardless of the different ejecta distribution on the antipodal hemisphere, the thicknesses at the antipode (within 1 degree) are consistent ($\sim$ 700 m) among the non-rotating case and the impacts at the poles (Figures \ref{fig:non45degree}C, \ref{fig:rot}A, and \ref{fig:rot}C).
This is expected since the location of the antipode is not affected by rotation for polar impacts.
Therefore, rotation of the Moon has a limited effect on the distribution of ejecta very close to the antipode for impacts occurring near the poles. 

\begin{figure}
\noindent\includegraphics[width=\textwidth]{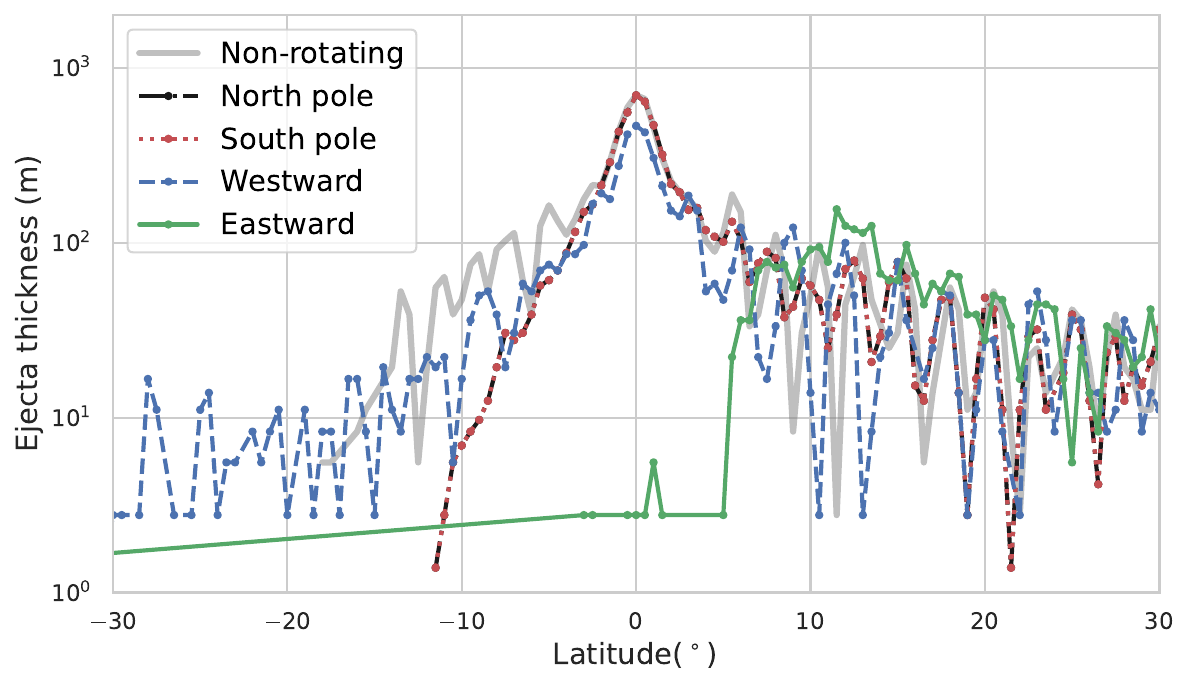}
\caption{Thickness of antipodal ejecta along the impact direction.
0$^\circ$ depicts the antipode, negative number denotes the direction over the antipode, and positive number denotes the opposite direction (between the antipode and the impact site). 
Each line represents a different impactor direction and impact site (see legend), and the gray line indicates the non-rotating case. 
Note that these are ejecta from the impactor, corresponding to panels A, C, E, and G in Figure \ref{fig:rot}.
}
\label{fig:rot_along}
\end{figure}

In contrast, rotation has a considerable effect on antipodal deposits for impacts occurring at the equator.
A westward impact at the equator produces a dispersed ejecta distribution, distributing more ejecta over the antipode. 
For the westward impact, the Moon rotates toward the point of impact as ejecta is in flight meaning ejecta that would usually be emplaced short of the antipode can now reach the antipode.
Because the flight time near and past the antipode tends to be longer (Figure \ref{fig:non45degree}A), the antipodal ejecta decreases to 70\% of the non-rotating case (Figure \ref{fig:rot}E). 
For an eastward impact, the Moon rotates away from the point of impact as ejecta is in flight meaning ejecta must travel a farther distance to reach the antipode. 
An eastward impact at the equator results in almost no antipodal ejecta (Figure \ref{fig:rot}G). 
Note that the antipodal thickness of 3 m corresponds to only two tracer particles, which is too small to be statistically reliable.
In this case, the accumulated deposition on the antipodal hemisphere shifts from the antipode, $\sim 12^\circ$ towards the impact location (Figure \ref{fig:rot_along}). 
These results indicate that consideration of the lunar rotation is crucial for ejecta from lunar basins near the equator.

\begin{figure}
\noindent\includegraphics[width=\textwidth]{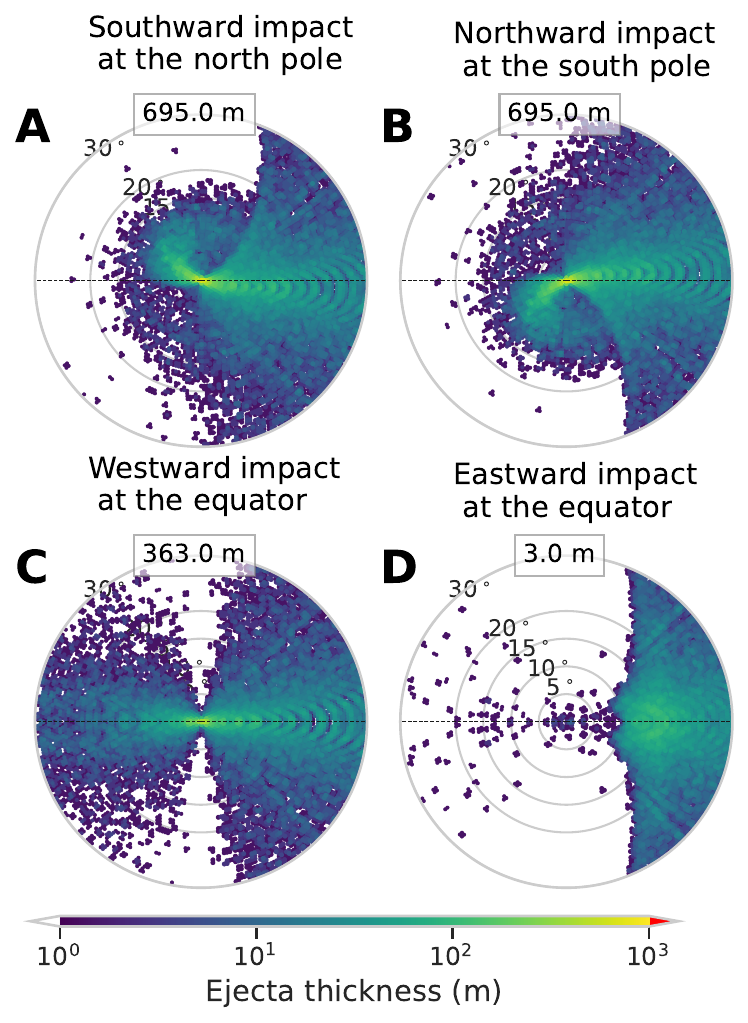}
\caption{Same viewing as Figure \ref{fig:rot}, but assuming the Moon has a rotational period of 17 days \cite{Daher:2021}.
This displays impactor ejecta only.
Note that these are shown in an antipode focused coordinate system (see Section \ref{sec:results} and Figure \ref{fig:rot}).
}
\label{fig:rot_17d}
\end{figure}

\begin{figure}
\noindent\includegraphics[width=\textwidth]{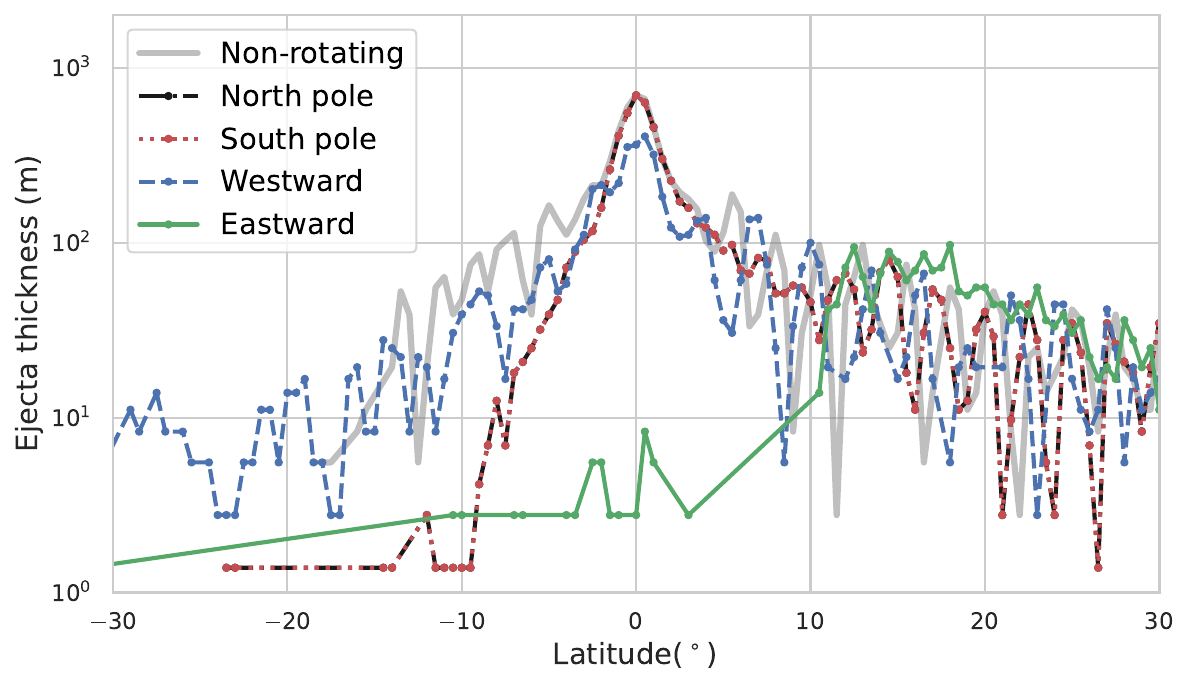}
\caption{Same viewing as Figure \ref{fig:rot_along}, but assuming the Moon has a rotational period of 17 days \cite{Daher:2021}.}
\label{fig:rot_along_17d}
\end{figure}

A faster rotation of the Moon further decreases the antipodal ejecta thickness. 
We also investigate the rotational period of the Moon as 17 days corresponding to the ancient Moon just after formation \cite<4.5 Ga,>{Daher:2021}.
The antipodal ejecta from the impact at the poles become more dispersed but look similar to our fiducial case of 27 days (Figures \ref{fig:rot_17d}A, \ref{fig:rot_17d}B). 
Their thicknesses at the antipode remain the same ($\sim$ 700 m) as the antipodes location is not affected by rotation for a truly polar impact. 
The westward impact at the equator results in more scattered ejecta, and the thickness at the antipode becomes about half of the non-rotating case ($\sim$ 360m, Figure \ref{fig:rot_17d}C). 
For the eastward impact at the equator, there are less ejecta at the antipodal hemisphere (Figure \ref{fig:rot_17d}D). 
The accumulated deposition approaches near to the impact location away from the antipode (Figure \ref{fig:rot_along_17d}).
Clearly the rotation period of the Moon can have a substantial effect on the distribution of antipodal ejecta, especially for impacts occurring closer to the equator. 

\begin{figure}
\noindent\includegraphics[width=\textwidth]{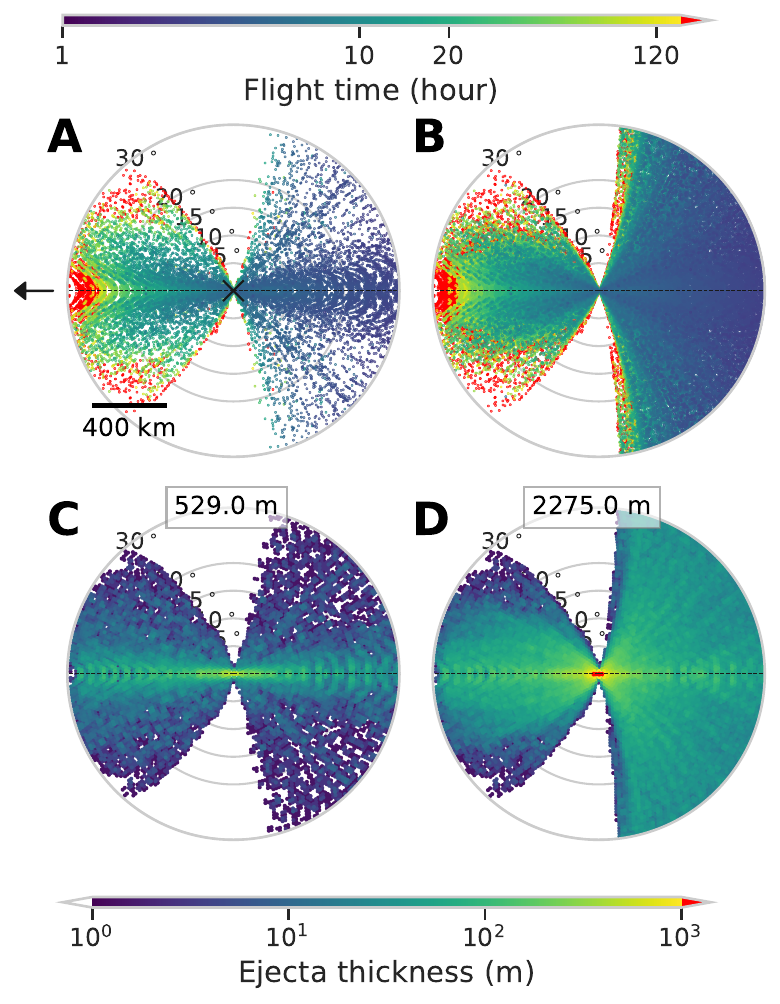}
\caption{
Same viewing as Figures \ref{fig:non45degree}, but the impact angle is 30$^\circ$.
Note that this is the non-rotating case.
}
\label{fig:non30degree}
\end{figure}

Since the obliquity of the impact affects the flight times and distribution of antipodal ejecta \cite{Wakita:2021}, it follows that the impact angle will also alter the effect of rotation on antipodal ejecta distributions . 
Figures \ref{fig:30degree} and \ref{fig:60degree} summarize the results of for impact angles of 30$^\circ$ and 60$^\circ$. 
The trends of antipodal deposition are similar for impact at poles, regardless of impact angle (Figures \ref{fig:rot}A, \ref{fig:rot}C, \ref{fig:30degree}A, \ref{fig:30degree}C, \ref{fig:60degree}A, and \ref{fig:60degree}C);
their distribution differ, but the thickness at the antipode is the same as the non-rotating case, as expected. 
Note that the abundant late arriving ejecta past the point of impact in 30$^\circ$ cases results in a sweeping \enquote{spiral arm} like pattern for the polar impacts.
However, antipodal depositions from impacts at the equator change significantly depending on the impact angle.
As some ejecta beyond the antipode by a 30$^\circ$ impact have flight times of less than 20 hours (Figure \ref{fig:non30degree}A), its path becomes shorter and results in accumulating at the antipode in the case of eastward impact at the equator. 
The antipodal ejecta thickness becomes two times thicker than the non-rotating case (Figures \ref{fig:non30degree}C, \ref{fig:30degree}G).
On the other hand, while a 60$^\circ$ impact has no antipodal ejecta without a rotation (Figure \ref{fig:non60degree}C), a westward impact scatters ejecta towards the antipode, making a 150 m thick ejecta deposition (Figure \ref{fig:60degree}E).
As such, even for the same impact direction, lunar rotation can result in either thickening or thinning of the antipodal ejecta depending on impact angle.

We also examine the rotational effect on the total ejecta at the antipode. 
Their behaviors are similar to the impactor ejecta (Figs. \ref{fig:rot}, \ref{fig:30degree}, and \ref{fig:60degree}), because their flight times are close \cite<see also>{Wakita:2021}. 
Additionally, as the total ejecta is more abundant than the impactor ejecta, lunar rotation more broadly (almost uniformly for some cases) distributes the total ejecta at antipodal hemisphere, except for eastward impact at the equator. 
This could make it difficult to identify the antipodal ejecta, unless they have unique features that differ from other materials, namely impactor material.

\section{Discussion} \label{sec:discussion}
\begin{figure}
\noindent\includegraphics[width=0.6\textwidth]{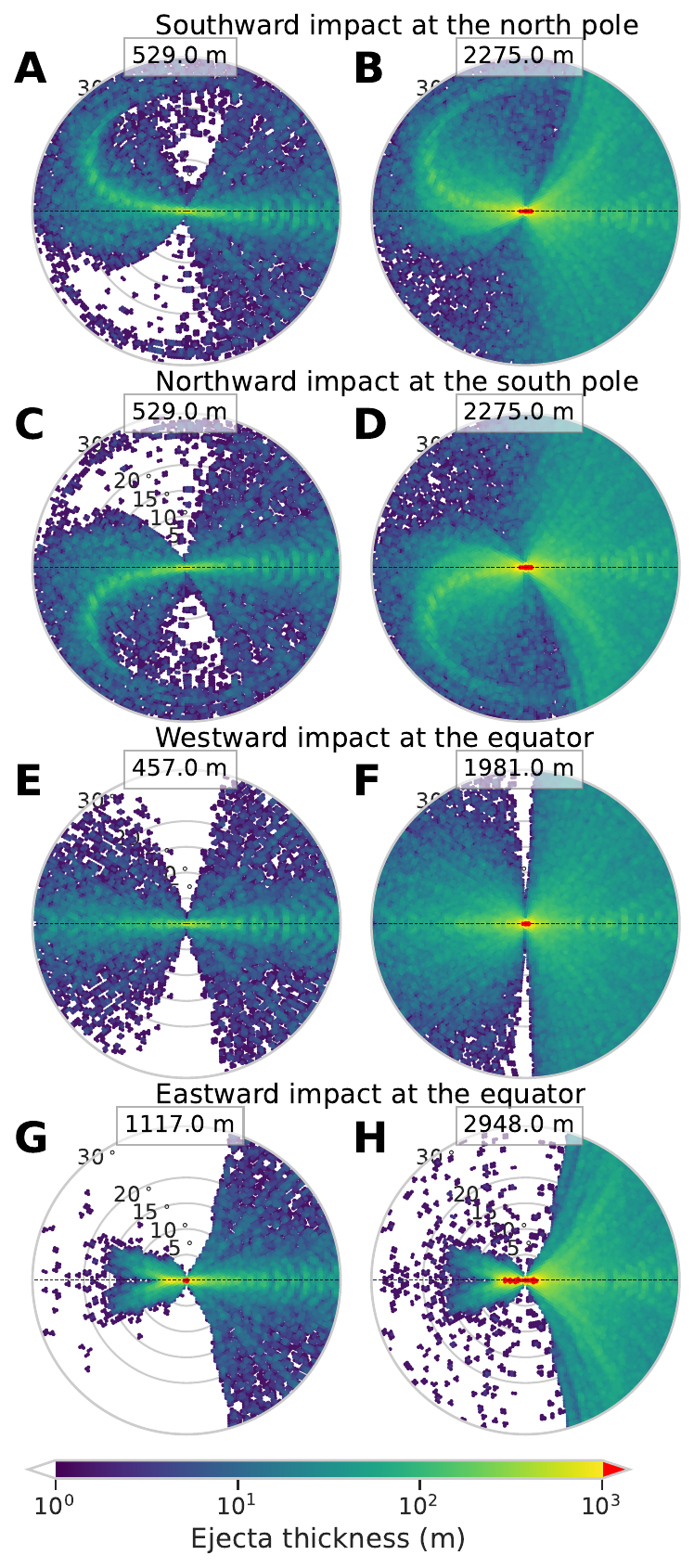}
\caption{Same viewing as Figures \ref{fig:rot}, but the impact angle is 30$^\circ$.
Note that these are shown in an antipode focused coordinate system (see Section \ref{sec:results} and Figure \ref{fig:rot}).
}
\label{fig:30degree}
\end{figure}

\begin{figure}
\noindent\includegraphics[width=0.6\textwidth]{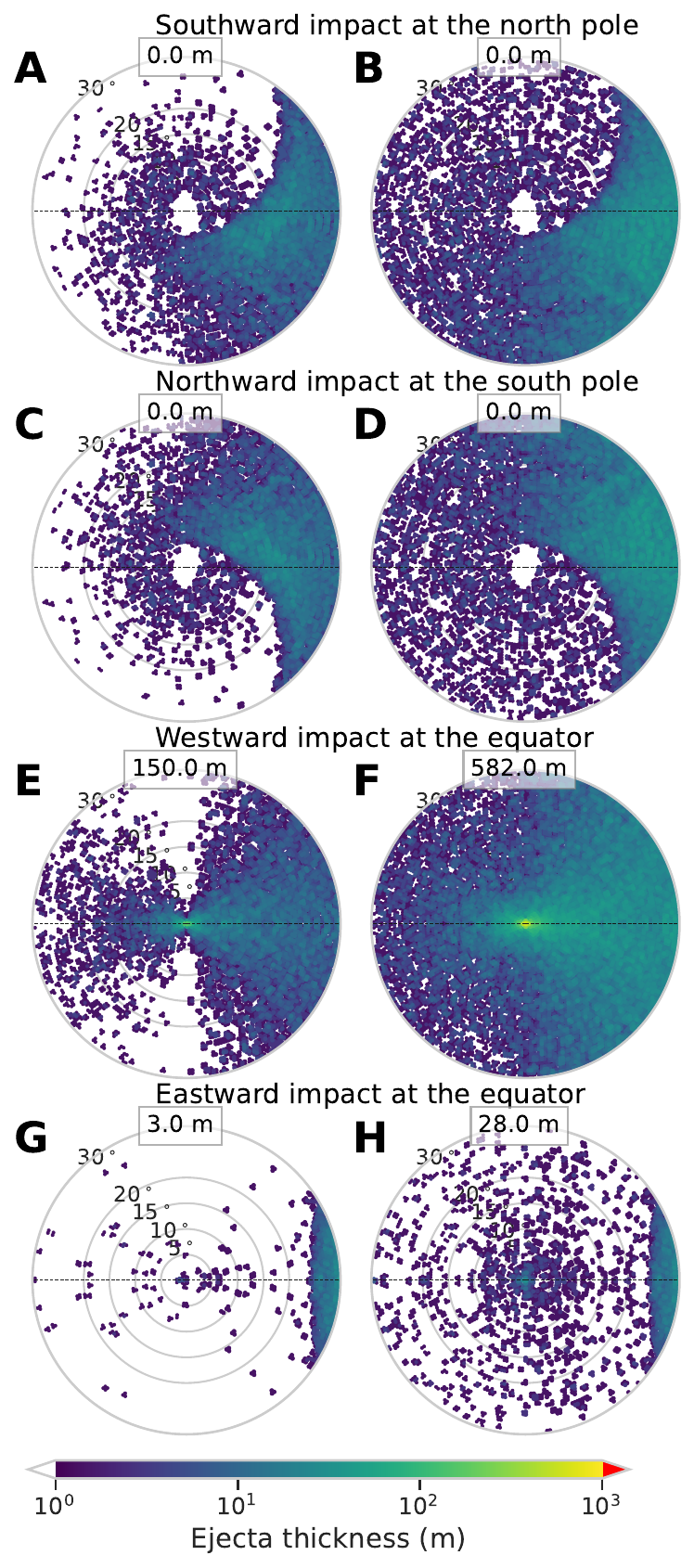}
\caption{Same viewing as Figures \ref{fig:30degree}, but the impact angle is 60$^\circ$.
Note that these are shown in an antipode focused coordinate system (see Section \ref{sec:results} and Figure \ref{fig:rot}).
}
\label{fig:60degree}
\end{figure}

Our results indicate that the effect of lunar rotation on the deposition of antipodal ejecta depends on the impact direction, location, and impact angle. 
For impacts near the poles, lunar rotation has a minor effect on the the ejecta thickness near the antipode (Figures \ref{fig:rot}A, \ref{fig:rot}C, \ref{fig:30degree}A, \ref{fig:30degree}C, \ref{fig:60degree}A, and \ref{fig:60degree}C). 
However, for impacts near the equator, the displacement of the antipodal ejecta is significant.
For an eastward impact at 45$^\circ$, there is almost no antipodal ejecta (Figure \ref{fig:rot}G) despite a thick deposit in the non-rotating case (Figure \ref{fig:non45degree}C). 
On the other hand, for an eastward impact at 30$^\circ$, antipodal ejecta becomes thicker than the non-rotating case (Figures \ref{fig:non30degree}C, \ref{fig:30degree}G). 
Moreover, there would be a thin antipodal ejecta deposit in the case of a westward impact at 60$^\circ$, even with no antipodal ejecta in the non-rotating case (Figures \ref{fig:non60degree}C, \ref{fig:60degree}E).
Our findings could give insight into the origin of lunar magnetic antipodal anomalies. 
Although antipodal ejecta from the impactor could record the lunar magnetic field, not all lunar basins near the equator possess antipodal magnetic anomalies.
If there are no magnetic anomalies antipodal to the basins, it may suggest that the impactor's direction and/or angle are unsuitable to accumulate the ejecta at the antipode. 
Therefore, we suggest that care must be taken in interpreting the magnetic anomalies antipodal to lunar basins.

Rotation of the Moon affects the landing location of impact ejecta, depending on their flight times and directions.
Although the eastward impacts at the equator are least effective in accumulating ejecta at the antipode, the accumulated ejecta thickness of $\sim 100$ m is thicker than other landing sites within 30$^\circ$ of the antipode (Figures \ref{fig:rot}G and \ref{fig:rot_along}).
We only investigate impact angles of 30, 45, and 60$^\circ$, however, other impact angles could also shift the accumulated deposition.
Since shallower angle impacts tend to produce more ejecta beyond the antipode, the ejecta could still accumulate around the antipode even when the eastward impact occurs near the equator (e.g., Figure \ref{fig:30degree}G). 
In contrast, although steeper angle impacts have limited or no antipodal ejecta \cite{Wakita:2021}, the rotation might broadly distribute them, resulting in making nearly antipodal ejecta deposits (e.g., Figure \ref{fig:60degree}E).
Even if there is less or no antipodal ejecta, the highly susceptible impactor material that landed in other locations could also record the ancient magnetic fields.
When ejecta have relatively short flight times, rotation of the Moon has less effect on their landing site. 
Those ejecta can explain the magnetic anomalies that lie on the path between the basin and its antipode \cite{Hood:2021}. 
These insights might help to decode the impact direction and angle of lunar basin forming impacts.

The faster rotational period of the ancient Moon further affects the antipodal deposition. 
While we investigate the lunar rotational period of 17 days \cite<4.5 Ga, >{Daher:2021}, we acknowledge that there could be other possibilities. 
The semi-major axis of the Moon at 4.5 Ga, which influences the ancient rotation period, could range from 30 Earth radii to 45 Earth radii \cite{Cuk:2019, Daher:2021}. 
The former corresponds to the rotational period of 10 days \cite{Cuk:2019}. 
Recent work by \citeA{Nimmo:2024a} suggests the Moon might be closer to the Earh as 19 Earth radii as late as 4.35 Ga.  
In such a faster rotating case, the effect of lunar rotation on antipodal deposition becomes more significant. 
Nonetheless, determining the lunar crater age and the rotational period simultaneously is difficult (see below).

We note that it is hard to determine the crater-forming impactor's direction, location, and angle. 
The impactor's direction could be decodable only if a crater has an ellipsoidal shape and/or an asymmetric ejecta blanket. 
It is suggested that the South Pole-Aitken basin could have formed via a southward impactor with 30$^\circ$ \cite<e.g.,>{Andrews-Hanna:2025,Wakita:2026a} with its antipodal ejecta deposited on the northern hemisphere. 
We also note that the observed basin center might differ from the impact site of the oblique impact. 
The basin center is likely to be located in the downrange direction from the impact location of the oblique impactor \cite{Schultz:2011,Elbeshausen:2009,Davison:2022}. 
Magnetic anomalies antipodal to the Crisium basin are offset a few degrees south from the Crisium's antipode \cite{Wakita:2021}. 
There are several possible explanations for this offset: the effect of lunar rotation due to the impactor's direction (e.g., south and west/east), the offset between the basin center and the impact site, and/or heterogeneous distribution of ejecta deposition. 
Moreover, the rotation axis of the Moon has been reoriented from its past pole location, a phenomenon known as true polar wander.
Observations suggested that the Moon's pole location may have moved by $\sim$6$^\circ$ to $\sim$35$^\circ$ \cite{Garrick-Bethell:2014,Keane:2014,Siegler:2016}. 
Furthermore, the basin-forming impacts themselves could temporarily unlock the Moon from synchronous rotation \cite{Melosh:1975,Wieczorek:2009,LeBars:2011}. 
This would further complicate estimating the antipodal deposition from basin-forming impacts. 
As such, we still need more investigation to apply our results to the observed lunar basins and their antipodal ejecta. 

We would like to mention the possibility of identifying lunar antipodal ejecta. 
One possible way to identify antipodal ejecta is through magnetic anomalies.  
Exogenous material tends to have higher thermoremanent magnetization susceptibility than a lunar material \cite<e.g.,>{Wieczorek:2012}. 
While the impact-induced shock would heat impactor materials above the Curie temperature, they would land at the antipode about 10 hour after the impact \cite{Wakita:2021}. 
Note that most antipodal ejecta from these lower velocity impacts remains unmelted (see Section \ref{sec:methods}).
Given that a 100 km diameter impactor will produce fragments of 50--1000 m \cite{Wiggins:2021}, 
they would be large enough to remain warm upon the landing and become magnetized as they cool. 
Thus, antipodal impactor material could record the magnetic field and exhibit magnetic anomalies, if the Moon had a magnetic field at the time of the crater-forming impact. 
While the impact direction and angles matter, our results imply that craters around the polar regions are more likely to have magnetic anomalies at their antipodes than the craters around the equatorial regions. 
Indeed, the magnetic anomalies around the north pole could correlate with craters at the south pole \cite<i.e., Schr\"{o}dinger;>{Hood:2013,Hood:2022}. 
Some magnetic anomalies near the equatorial region are antipodal to lunar basins \cite<e.g., Orientale, Crisium;>{Hood:2013,Hood:2021}.
According to our results, these may be produced by eastward impacts with shallower impact angles ($\le 30^{\circ}$) or westward impacts occurring at steeper angles ($\ge 60^{\circ}$). 
As discussed above, future work is necessary to confirm whether antipodal ejecta caused these magnetic anomalies. 
Investigations into magnetic anomaly offsets and/or unique ejecta distributions caused by lunar rotation (e.g., \enquote{spiral arm} like pattern) could also help identify paleo-impact directions and locations, which may have shifted due to true polar wander.
Differentiated impactors that have an iron core could also result in producing additional magnetic anomalies near the impact point \cite{Wieczorek:2012,Citron:2025}. 
Note that the impact-induced shock would also result in the antipodal focusing \cite<e.g.,>{Schultz:1975}. 
The acquisition of a shock remanent magnetization at the antipodal crust would precede the antipodal ejecta’s landing \cite<about 1 hour after the impact, see >{Narrett:2025}. 
We, however, note that the impact ejecta are unlikley to acquire a shock remanent magnetization because of the low shock pressure upon landing \cite{Garrick-Bethell:2020}.

Antipodal ejecta could also be observable as compositional or topographic anomalies \cite<e.g.,>{Wieczorek:2001}. 
Exogenic impactor material could be compositionally different from lunar material. 
While their accumulation at the antipode may be detectable as compositional anomalies, later mixing with the target material could suppress their appearance.
In addition, the antipodal ejecta from the target likely have a longer flight time than the impactor ejecta \cite{Wakita:2021}.
Although the mixing process could occur upon the landing, the target ejecta could potentially cover the earlier deposition of the impactor antipodal ejecta by up to $\sim$2000 m (see Figs. \ref{fig:rot}, \ref{fig:30degree}, and \ref{fig:60degree}), reducing the detectability of compositional anomalies from the impactor. 
On the other hand, the target material could be ejected from a depth of up to 20 km \cite{Wakita:2021}. 
Since they could be compositionally different from material at the antipode, compositional anomalies might still be observable. 
The accumulation of the antipodal ejecta may appear as topographic anomalies, regardless of their composition. 
However, these could be observable only if the eastward impact occurs near the equator. 
In other cases, the total ejecta is distributed over most of the Moon and exhibits a weak antipodal focusing, which would inhibit formation of detectable topographic mounds or compositional differences (see Figs. \ref{fig:rot}, \ref{fig:30degree}, and \ref{fig:60degree}).

\section{Conclusion} \label{sec:conclusions}
We explore the effect of the Moon's rotation on antipodal ejecta deposition. 
An impact at the poles distributes antipodal ejecta more broadly than the non-rotating case, but it still produces the same thickness of antipodal ejecta as the non-rotating case.
In contrast, lunar rotation considerably changes the antipodal ejecta if impacts occurs at the equator.
In our fiducial case of impact angle of 45$^\circ$, while a westward impact near the equator still produces antipodal ejecta, an eastward impact case has almost no ejecta at the antipode. 
The eastward impact exhibits enhanced deposition towards the impact location away from the antipode, and it's still thicker than other landing locations. 
However, since an impact at 30$^\circ$ angle produces more ejecta beyond the antipode, an eastward impact results in a thicker antipodal deposit even compared to the non-rotating case.
Additionally, a westward impact at 60$^\circ$ incidence produces an antipodal ejecta deposition, while there is none in the non-rotating case.
The effect of the lunar rotation period on the antipodal ejecta deposit is also significant for impacts at the equator. 
These results indicate that consideration of lunar rotation must be taken for ejecta from lunar basins, especially for those located near the equator.
Thus, we suggest that particular care be taken when interpreting magnetic anomalies found antipodal to lunar basins.



%
%

%
%
%
%
%
%
%
%


\clearpage
\section*{Open Research Section}
Our outputs are available in \citeA{Wakita:2025c}.

\acknowledgments
This work was supported by NASA SSERVI Grant 80NSSC23M0161.
This research was supported in part through computational resources provided by Information Technology at Purdue, West Lafayette, Indiana.
We gratefully acknowledge Mark Wieczorek for fruitful conversations that spurred this study.
We thank the reviewers for their helpful feedback, which improved this manuscript.

\section*{Conflict of Interest}
The authors declare no conflicts of interest relevant to this study.

%
%


%
%
%
%
%

\end{document}